# Thermal Pressures at the melt based on first principles derived for the transition metals: iron (3d), chromium (3d), vanadium (4d), iridium and platinum (5d)

Joseph Gal*

Ilse Katz Institute for Nanoscale Science and Technology ,
Ben-Gurion University of the Negev, Beer Sheva ,84105 Israel

*jgal@bgu.ac.il

## Abstract

The pressure reported by diamond anvil cell (DAC) and shock waves (SW) experiments in most cases do not match and hardly represent the actual pressure experienced by the examined sample. Recently, Y. Zhang et al. proposed correction procedure of the reported SW data for elemental vanadium. This approach was strongly supported by the first principles DFT-Z method, proving that this procedure is indeed reasonable. It is suggested that this procedure should also be applied to other d-electrons the transition metals Fe, Ir and Pt. Thus, the actual pressure-temperature scales in DAC and SW experiments can be corrected by taking in to account the thermal pressure shifts. In the present contribution it is further claimed that first principles *ab-initio* DFT and MD simulations should serve as an anchor for correcting the pressures and temperatures reported by DAC and SW experiments. It is concluded that upon deriving the actual pressure sensed by the explored sample, the thermal pressure and the temperature shifts must be taken into account when constructing melting curves. In addition, the advantage of the Lindemann-Gilvarry vs. Simon-Glatzel fitting procedure of melting curves is claimed.

# 1. Introduction

During the last six decades, diamond anvil cells (DAC) has been frequently used to determine the equations of state (EOS) and melting curves of elemental metals [1]. However, melting curves derived by DAC experiments have never matched the shock waves (SW) experimental results nor theoretical density functional theory (DFT) simulations [1,2,3,4] where chromium metal is an exception (see discussion). It is well in consensus that the pressure transmitting medium (PTM) and the packing procedure of the investigated sample, namely the packing environment and the gasket type, plays a role in the inconsistencies of reported melting curves. This rise the question whether DAC measurements contribute to better understanding of the high pressure physics.

Many experiments have shown that the melting curves of the same element measured by a laser-heated DAC (LH-DAC), reported by different experiments, reveal different melting points using different PTMs [5,6]. The pressure medium is expected to distribute the pressure homogeneously within the pressure chamber, preventing non-hydrostatic effects such as, pressure gradients, shear stress, or inhomogeneous pressure [1]. Upon increasing the temperature at each pressure the examined sample and the PTM both are subject to increases of their volumes. Nevertheless, the volume expansion is suppressed by the chamber's finite volume, provoking an increase in the thermal pressure over the whole system. In some cases, the PTM melts and remains liquid during the entire experiment [7]. Therefore, the pressure reported does not necessarily present the actual pressure experienced by the sample. The effects of the PTM on the measured sample, and the PTM's response to P,T changes throughout the experiment must be taken into account. Therefore, the pressure scale of the reported melting curves and the equations of state (EOS) isotherms should be corrected according to the actual pressure in the cell and illustrative examples are shown below.

In a recent publication, Y. Zhang et al. [8] took into account the radiation absorption by the LiF window; thus correcting the reported SW data for elemental vanadium. This approach is strongly supported by the first principles DFT-Z method, proving that this procedure is indeed reasonable. In addition, Y. Zhang et al. suggested that this procedure should also be applied to all the d-electron transition metals. Zhang's results clearly confirm the linear increase of the thermal pressure shifts upon elevating the applied temperature (see Figs 1-4) as predicted by first principles theories [9].

In a previous paper, it was argued that if isochoric conditions exist in the DAC chambers [4], thus increasing the temperature provokes an additional thermal pressure. In practice, a mechanical pressure gauge to measure this thermal pressure ($P_{th}$) does not exist. Therefore, the actual pressure must be estimated, either from the pressure shift of each melting point relative to the initial ambient pressure (see Figs 1-4), or from first principles calculations of the lattice component in the P-V-T equation of state [9]. The direct determination of the thermal pressure and temperature at the melt from experimental results was not available up to date. In the present contribution, the linear rise of $P_{th}$ vs. the temperature, as predicted by first principles assumptions [9] is confirmed. This allows the extrapolation of the thermal pressure and the thermal temperature up to the melt, thus correcting the pressure scales and the melting temperatures of the transition metals Fe, V, Ir and Pt.

## 2. Analysis of the published data.

All the presented data in the present paper were downloaded and reanalyzed using *B. Tummers,* datathief [*10*]. The obtained melting curves were fitted by applying the constraint that the fitting volume parameter V of the experimental equation of state (EOS) at ambient temperature must simultaneously fit the experimental melting and the EOS data. Note that the Grüneisen parameter $\gamma$ is a fitting parameter assigned in the manuscript "combined approach" [see Appendix]. The corroboration of this fitting procedure is confirmed by the *ab-initio* DFT- Z methodology [3].

## 3. Melting curve of elemental vanadium

The melting curve of vanadium reported by Y. Zhang et al. [8] (therein Fig.4) was determined by in-situ x-ray diffraction in a LH-DAC and is depicted in Fig. 1. The blue solid line presents the melting curve of V fitted with the combined approach constraint, which is confirmed by the first principles simulations (red solid line, DFT-Z method [3]) . In Fig.1(**a**), the colored asterisks represent the pressure thermal shifts as a function of the applied temperature. The dashed black lines represent the 300K applied pressures. The angle (difference) between the dashed lines and the asterisk points, indicate that the thermal pressure ($P_{th}$) is provoked by increasing the temperature, thus confirming the assumption that  isochoric conditions exist in the DAC chamber when using KCl PTM. The horizontal colored double arrow lines indicate the pressure thermal shifts ($P_{th}$) at the melt. In Fig.1(**b**) the corrected melting curve of V as derived from Fig.1(**a**) is presented**.**

The contribution of Y. Zhang at al. [8] clearly demonstrates that all the experimental SW and the DAC data reported to-date, need pressure and temperature scales corrections. As shown, there is no discrepancy between the static DAC and the dynamic SW measurements of V, if the radiation absorption by the LiF window is taken correctly into account. The proposed approach is indeed confirmed by first principles, using the *ab-initio* DFT-Z method simulations [3].

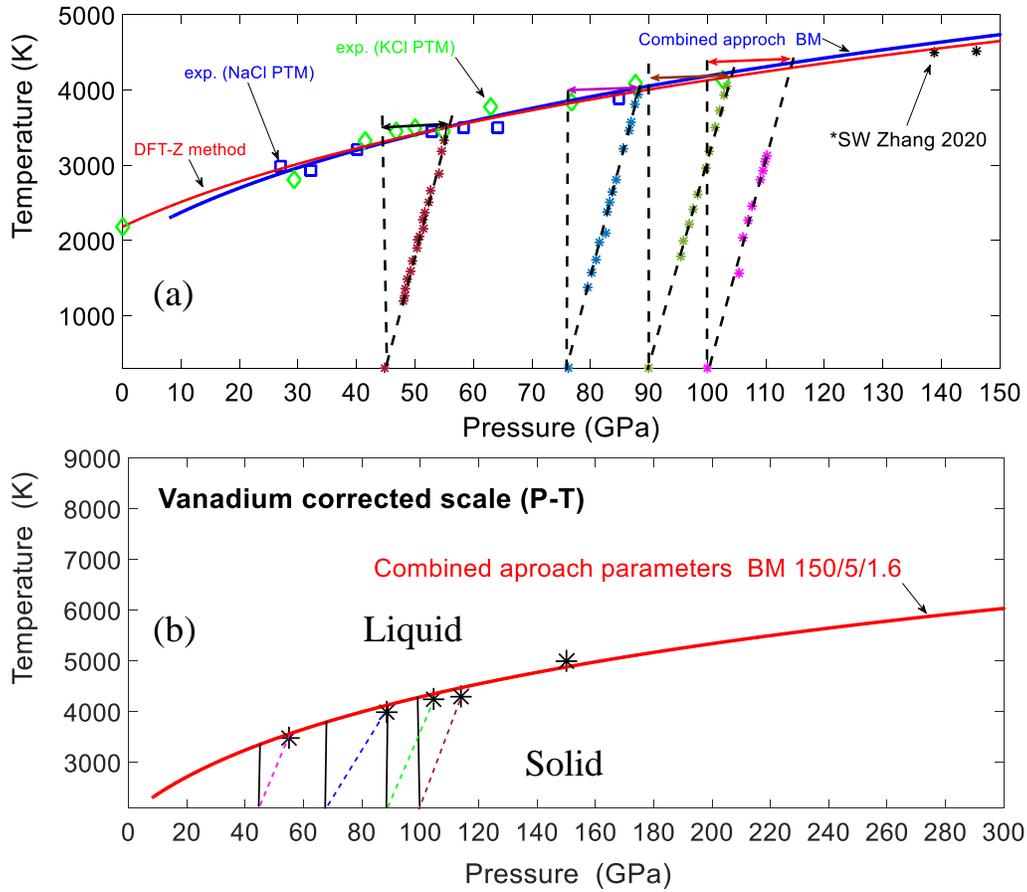

**Fig. 1(a):** Thermal pressure shifts at several applied pressures of vanadium metal. The red and blue solid lines represent the melting curve of V fitted with the combined approach approximation with BM parameters (150/5/1.6) and further is confirmed by DFT-Z method simulation. The blue squares are the experimental melting points derived LH-DAC with V embedded in a liquid NaCl PTM [6]. The green diamonds represent V embedded in a solid KCl PTM. The 100 GPa melting point is V embedded in a MgO PTM. The colored asterisks are the thermal pressure shifts as a function of the applied temperature, reported by Y. Zhang [8]. The dashed black lines present the ambient applied pressures. The angle between the dashed lines and the asterisk points, indicates the thermal pressure ($P_{th}$) shifts provoked by the increase of the temperature. The horizontal double arrows indicate the thermal shifts ($P_{th}$).
**(b)**: Proposed corrected scale of V melting curve. The colored dashed line represent the pressure-temperature shifts which are derived from Fig.1**(a)**.

## ε - Iron

Isochoric behavior of ε - Iron was reported by R. Sinmyo et al. (2019) [11] and is depicted in Fig.2(**a**). This result totally contradicts the quasi-isobaric behavior of ε iron in laser heated DAC claimed by Anzellini et al. [12]. The reason of this discrepancy comes from the packing of the iron foil with $Al_2O_3$ adding Ar PTM compared to KCl PTM in Anzellini's experiments.

Four melting points of iron ε-phase (out of 10) reported by R. Sinmyo et al. [11] are displayed in Fig.2(a), corroborating the melting points claimed by C. Murphy et al. (2011), where the thermally corrected melting points (marked X) are derived by inelastic X-ray scattering and phonon density of states [13]. The green squares in Fig.2(a) are the melting data reported by Anzellini et al. [12], therein Fig.2. The red open circles are DAC melting results derived by Sinmyo et al., fitted by the combined approach constraint (green solid line). The purple open circle is the triple point liquid-fcc-hcp structures. The colored asterisks represent pressure-temperature thermal shifts confirming isochoric condition in the DAC chamber. The horizontal colored double arrow lines indicate the pressure thermal shifts ($P_{th}$) at the melt, based on the theoretical *ab-initio* melting curve. The black O are the experimental results obtained by Tateno et al. [14] indicating that iron at pressures above the ICB is in the solid state.

The pressure – temperature corrected scale based on the theoretical *ab-initio* anchor deduced from Fig.2(a), is the main result for the melting curve of ε-iron metal. Note that the red asterisks in Fig.2(b) show the $P_{th}$ shifts at the melt which are fitted by the combined approach with Vinet fit parameters 163.4/5. 55/1.6 (blue solid line, see appendix). Possible explanation of the discrepancy between Anzellini et al. [12] and Sinmyo et al. [11] experiments is given in the discussion.

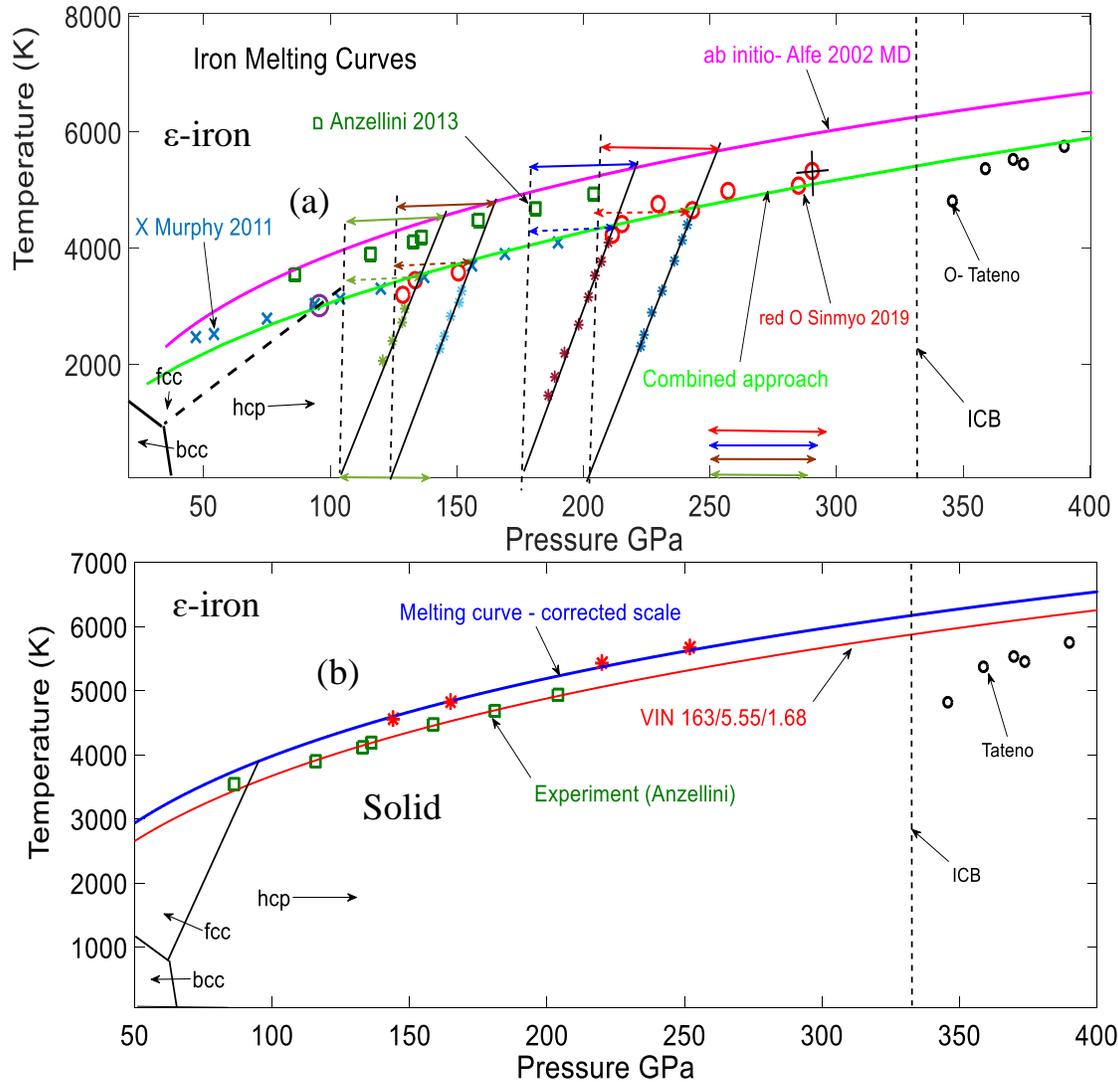

**Fig.2**: Melting curve of ε-iron. **(a)**: The magenta solid line represents the theoretical molecular dynamics (MD) ab-intio calculated melting curve Alfe et al. [15]. The double arrows lines indicate the thermal shifts relative to the theoretical anchor and the experimental melting curve (dashed). The green squares present the experimental DAC melting points reported by Anzellini et al. [12]. The red open circles are DAC melting results derived by Sinmyo et al. [11] fitted by adopting the combined approach constraint (green solid line). The purple open circle is the triple point. The blue X points are C. Murphy et al. [13] thermally corrected melting points, derived by inelastic X-ray

scattering The colored asterisks represent pressure-temperature thermal shifts as reported R. Sinmyo et al. [11]. The black O's present Tateno et al. experimental results [14]. **(b)**: Pressure – temperature corrected scale based on the theoretical ab-initio anchor, deduced from Fig.2**(a)**, marked by the red asterisks, fitted by the combined approach with Vinet fit parameters 163.4/5. 55/1.6 (blue solid line, see appendix).

# Iridium

Iridium attracted considerable interest in the scientific community due to its outstanding elastic and thermal properties. It is one of the most incompressible metal in nature, exhibiting face-centered cubic (fcc) structure up to 78 GPa with measured bulk modulus according to the combined approach calculation $B_o$= 341(3) GPa. Based on *ab-initio* simulations, above ~78GPa up to 600 GPa the r-hcp structure was proposed [3]. At room temperature (RT) Ir exhibits a density of 22.56 g/cc, and a high shear modulus, $G_o$ = 210 GPa. Thus, thanks to its shear modulus, chemical inertness, refractory nature, and phase stability, Ir is used for crucibles and thermocouples. The experimental pressure – temperature data at 26 and 36 GPa of fcc phase derived by Anzellini et al. [16] is depicted in Fig.3(a). The thermally corrected scale is depicted in Fig.3(b). In addition, as shown in Fig.3(c) a triple point is expected at ~ 80 GPa (500K) [3] predicting a phase transition from fcc to a disordered hexagonal close-packed phase (r-hcp). The extrapolation of the combined approach up to 600 GPa (red solid line) with the same fitting parameters 341/4.4/2.55, is corroborated by the DFT simulation. The dashed black lines present the ambient applied pressures. The angle between the dashed black lines and the colored round points, are due to the thermal pressure ($P_{th}$) shifts provoked by the increase of the temperature. The horizontal black double arrows in Fig.3(c) indicate the thermal pressure shifts ($P_{th}$).

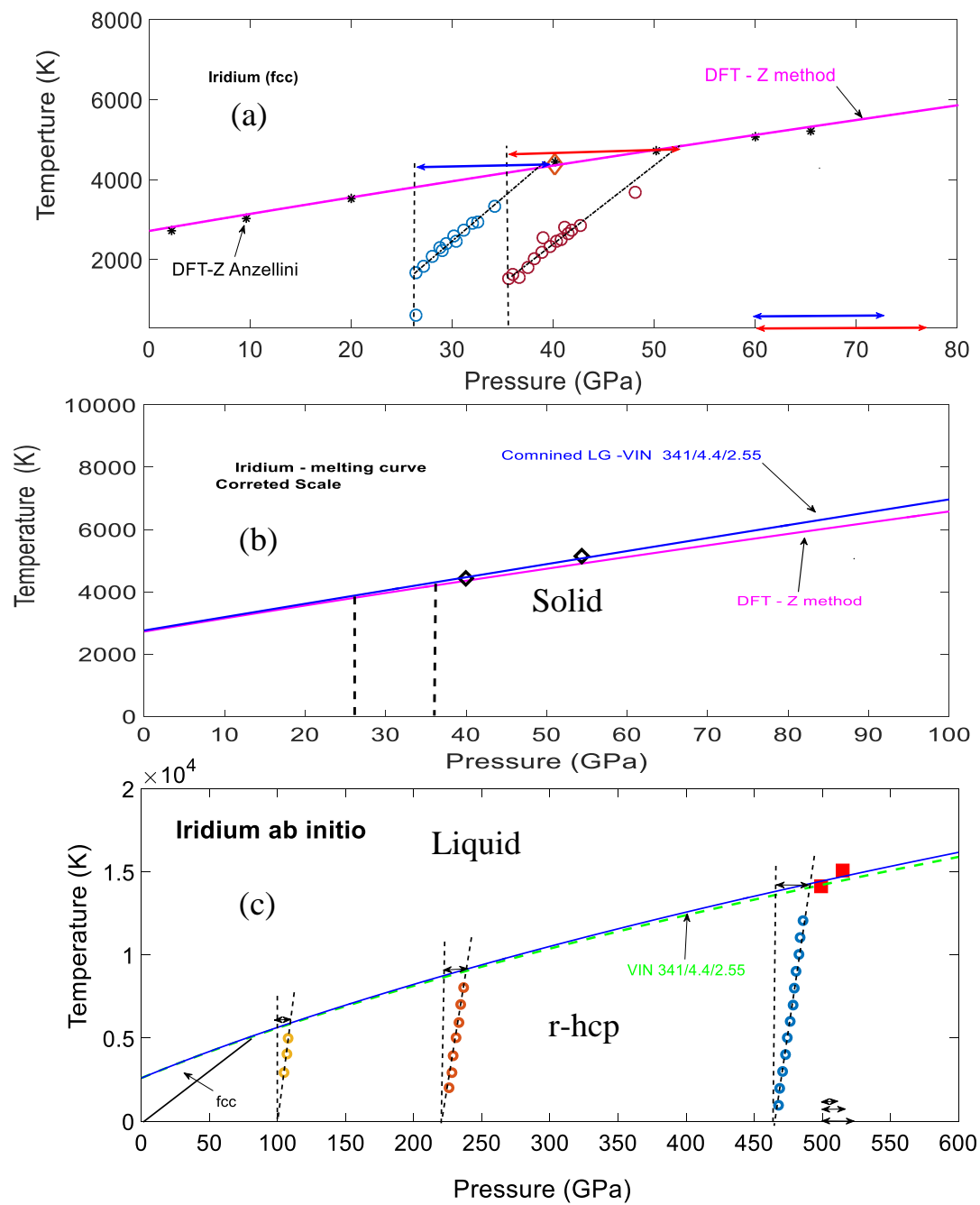

**Fig.3**: **(a)** Experimental pressure – temperature of Ir (fcc) phase at 26 and 36 GPa [16]. The melting curve is determined by theoretical DFT-Z method simulations [3]. The red and blue double arrows indicate the measured thermal shifts at the melt relative to the measured pressure at 300K, marked by the dashed lines. **(b)** Ir melting curve, pressure corrected scale derived from Fig.3(**a**). The blue solid line represents the constrained Lindemann- Gilvarry melting formula with Vinet EOS (combined approach) assigned in the figure ($B_o$,$B_o$', $\gamma_o$) 341/4.4/2.55 where $\gamma_o$ is a fitting free parameter. The dashed lines represent the initial pressure as measured experimentally at RT. The magenta solid line presents the DFT-Z simulation (L. Burakovsky [3]). (**c**): The colored open circles present the Z method calculations of the EOS (P-V) for r-hcp phase. The fcc phase is fitted by the third order Birch-Murnagham (BM) and Vinet (VIN) EOS. The extrapolation of the combined approach up to 600 GPa (red solid line) was performed with the same fitting parameters 341/4.4/2.55.

## Platinum

Platinum is interesting as it exhibits fcc to r-hcp phase transition upon increasing the pressure as shown in Fig.4(a).

Recent reported thermal shifts by Anzellini et al. [17] (therein Fig.6), a difference in the pressure shifts between the fcc and hcp phase is clearly recognized as demonstrated in the figure. The corrected pressure scale at the melt, derived from Fig.4(a) (colored double arrows), is proposed in Fig.4(b). The pressure-temperature melting points (blue diamonds) at the r-hcp region are shown in the figure. The corrected melting points are fitted with the combined approach (red solid line) which is corroborated by the first principles Z-method (green hexagrams) [3]. By utilizing the EOS parameters according to Zha et al. [18] (red solid line) the corrected melting points are fitted with BM 273/4.2/2.63 parameters where $\gamma_o$ is a free parameter.

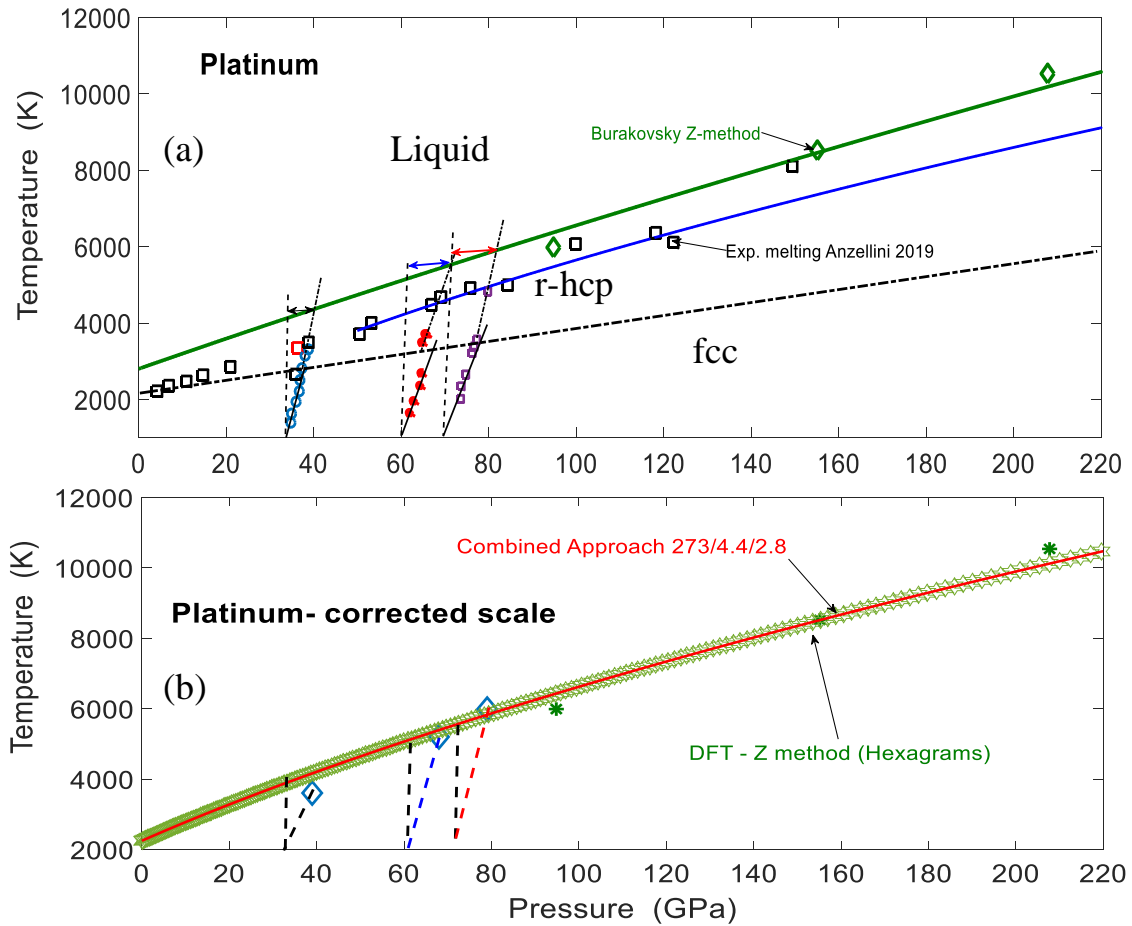

Solid

Fig.4 **(a):** Melting curve of platinum based on the DFT- Z method (solid green line). The black squares are the experimental melting points derived by DAC reported by Anzellini et al. (2019) [17]. The red square represents the triple point. The double arrows show the thermal shifts at the r-hcp structure region. Note the different shifts in the fcc region (black solid lines) compared to the shifts in r-hcp region dashed lines. The dashed black line presents the solid-solid phase boundaries. The solid blue line presents the fitting of the experimental data utilizing the combined approach with BM parameters proposed by Zha et al. 273/4.2/2.8 where $\gamma_o$ is free parameter. **(b)** The corrected pressure-temperature melting points (blue diamonds) at the r-hcp region. The corrected melting points are fitted with the combined approach (solid red line) corroborated by the first principles Z-method (green hexagrams) [3]. By utilizing the EOS parameters according to Zha et al. [18] (red solid line) the corrected melting points are fitted with BM 273/4.2/2.63 parameters where $\gamma_o$ is a free parameter.

# Discussion

It is assumed that the temperatures and the pressures of the present transition metals are indeed correct, and the derived temperatures either by a thermocouple or by optical pyrometer (black body spectra), within the errors, are reasonable [1]. The advantage or disadvantage of these methods are beyond the frame of the present contribution. However, the fact that the data, within the errors, of V DAC experiments imbedded in NaCl or KCl match the corrected SW data lead to the conclusion that the reported initial pressures and temperatures are reliable.

In many published articles isochoric or quasi-isobaric conditions in the DAC chambers have been reported [7,11,12,14]. Quasi-isobaric behavior means that due to temperature rise the sample expansion could not be suppressed by the PTM leading to melting curves lower than predicted by ab-initio simulations.

Nevertheless, in the case of isochoric condition in the DAC means that the PTM prevent thermal volume expansion thus provoking increase in the thermal pressure.

In conclusion the pressure transmitting medium (PTM) and packing procedure of the of the examined sample explain the discrepancies between the experimental reported melting points. In addition, in the case of V, by taking into account the absorption of the LiF window in SW experiments no discrepancy between SW and DAC results exist [8]. This lead to the conclusion that first principals DFT- Z methodology [3] should serve as an anchor for the pressure and temperature scale corrections.

The actual pressure on V sample in the LH-DAC chamber can be estimated from first principles calculations utilizing the P-V-T equation of state [9]. However, the use of experimental data to directly determine the actual pressure and temperature at the melt in a LH-DAC experiments has not been possible up to now. In the present contribution, a method to directly derive the thermal pressure and the temperature at the melt is proposed. The corrected scale (actual pressure) of V melting points and the proposed melting curve are depicted in Fig.1(b). Note that the experiment proposed by Zhang et al. was performed with KCl PTM [8], while the experiment reports by D. Errandonea et al. was performed with NaCl PTM [6]. The corrected SW data, which is confirmed by the DFT-Z method [3], is a proof that the correction proposed by Y. Zhang et al. is reasonable and can be applied to other transition metals like for example Fe, Ir or Pt as follows:

In the case of $\varepsilon$-iron metal, the colored asterisks in Fig.2(a) represent pressure-temperature thermal shifts as expected in isochoric condition in the DACs chamber. This isochoric behavior relate to the packing proposed by R. Sinmyo et al. ($Al_3O_3$+Ar). Again, the linear increase of the thermal temperature as theoretically predicted by the P-V-T equation of state [9] is experimentally confirmed.

Upon raising the temperature and pressure the volume expansion of iron metal, as reported by Anzellini et al. [12], relates to the fact that approaching the melt from the solid with KCl PTM, a mixture of the $\varepsilon,\beta,\gamma$ phases could exist. The existence of the $\gamma$ phase at high pressures and temperatures have been reported by Mikaeylushkin et al. [19] showing that the $\gamma$ phase can exist even at 165 GPa confirmed when quenched to room temperature. In addition, S.K. Saxena al al. [20] reported observation of the β phase above 110GPa and 3000K. Deep look in to Fig.1a of Anzellini et al. reveal that at 133GPa and 4292K (green dots their-in, [12]) at the angle 11.58(2Θ) there could exist a possible reflection which is ignored by the authors. This reflection could be analyzed as (111) gamma (γ) phase or dhcp (100) (β) phase. Thus, approaching the melt using KCl PTM from the solid iron a possible mixture of ε,β,γ phases can't be ruled out. This explains the discrepancy between Anzellini et al. [12] and Sinmyo et al. [11] experiments.

In the case of iridium, the extrapolation of the combined approach up to 600 GPa (Fig.3(c) red solid line) with the same fitting parameters for low and extreme pressures (341/4.4/2.55) corroborate the assumption of L. Burakovsky et al. that a second order smooth transition from fcc to hcp exist [3]. The colored open circles in Fig.4(c) present the Z method calculations of the EOS (P-V) for the fcc and r-hcp phases. The dashed black lines present the angle between the applied pressures and the calculated thermal shift ($P_{th}$), similar to V and Fe. Note that L. Burakovsky et al. relate the phase transition fcc to r-hcp to 5d electron rearrangement. The loss of the fcc stability drives the hex structure by a second order phase transition suggesting a random layer stacking.

The elastic properties of platinum show high density, low strength, usually used for crucibles at low temperatures. In addition, platinum serves as pressure standard in DAC X-ray experiments as well as in high-pressure shock experiments. There have been earlier reports of experimental melting curve for Pt [21,22]. The difference in the thermal pressure shifts between the fcc and hcp is clearly observed and is demonstrated in Fig.4(a), where the corrected melting scale at the melt (r-hcp structure) is depicted in Fig.4(b). Similar to ε-iron, the experimental results are below the first principles calculated melting curve. We attribute this behavior of the Pt metal to the KCL PTM in both experiments.

Nevertheless, a large uncertainty in the estimated temperature, fitted with Planck function in the grey-body approximation and Wien pyrometry, could lead to a large span in the errors of the reported temperatures in DAC measurements.

The phenomenon of isobaric behavior of vanadium recrystallization [6], different from the isochoric behavior claimed by Y. Zhang et al. [8] must be related to the elastic properties of NaCl PTM. If not, this calls for a theoretical explanation.

In Fig.5 high pressure studies of chromium metal utilizing KBr PTM is depicted. Cr metal is the only case where the theoretical approach, using ab initio quantum molecular dynamics (QMD) simulations based on the Z methodology match the DAC experiments [23,24]. It demonstrates the advantage of the combined Lindemann-Gilvarry approach vs. vs. Simon-Glatzel fitting curve procedure, revealing Bo, Bo' and $\gamma_o$.

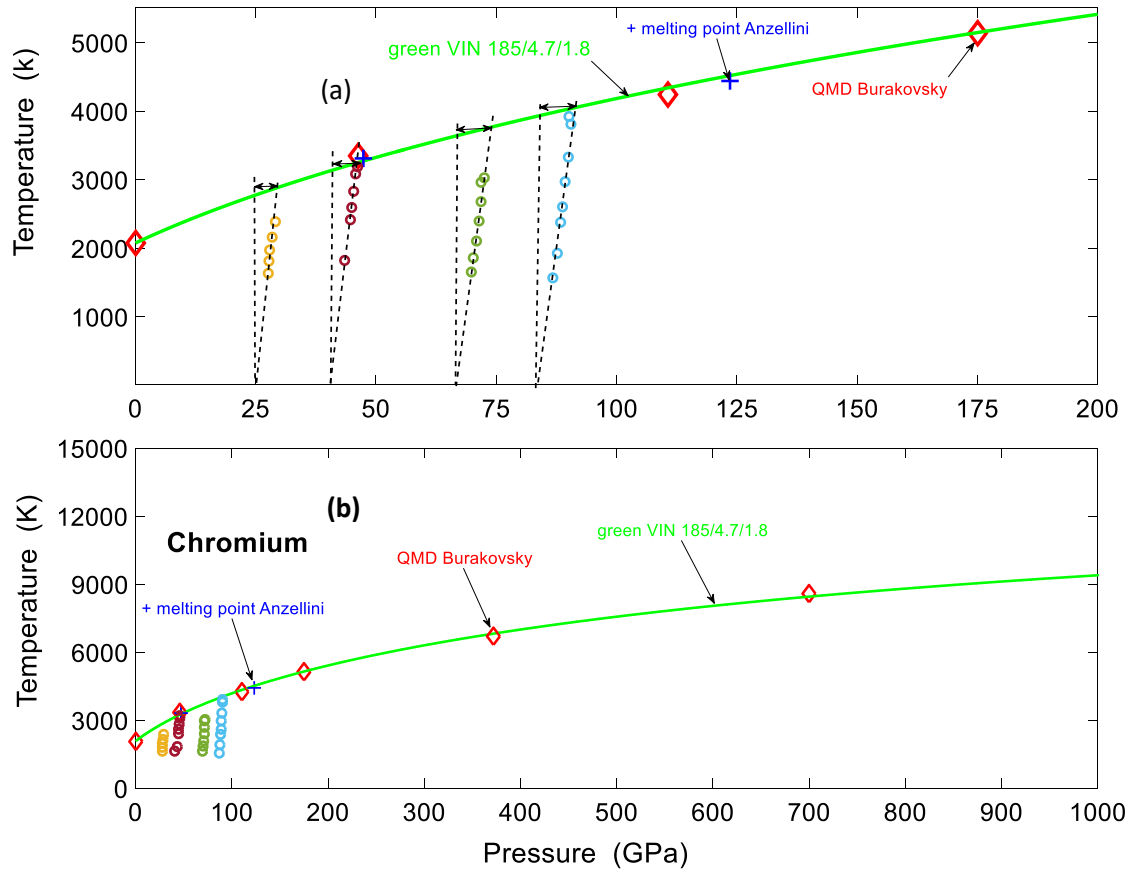

Fig.5 : Melting curve of Cr metal. (a): The colored circles are the experimental data measured by Anzellini et al. [23]. The double arrows show the thermal shifts, bcc structure region. The red diamonds represent S. R. Baty and L. Burakovsky et al. QMD-Z methodology calculations [24]. The green solid line represent the Lindemann- Gilvarry melting formula with Vinet EOS (combined approach) assigned in the figure ($B_o$, $B_o'$, $\gamma_o$)  185/4.7/1.8, where $\gamma_o$ is a fitting free parameter. (b) Calculated melting points extended to 1000 GPa.

## Conclusions

The pressure and temperature scales reported in DAC experiment do not represent the actual pressure experienced by the sample in the cell. The different response of the PTM's to P,T changes is the reason for the variety of melting curves reported in the literature.

It is suggested that the melting curves derived by the first principals DFT- Z methodology should serve as an anchor for the pressure corrections.

Upon deriving the actual pressure sensed by the explored sample, the thermal pressure and the temperature shifts must be taken into account when constructing melting curves.

The corrected pressure scales for metallic Fe (3d), V (4d), Ir and Pt (5d) transition metals are proposed for the first time.

## Acknowledgements

The author gratefully acknowledge Prof. Z. Zinamon, Department of Particle Physics, Weizmann Institute of Science, Rehovot – Israel, for the many helpful and illuminating discussions and comments.

# Appendix

1. **Gilvarry-Lindemann approximation:** The Combined Approach.

According to Lindenmann's criterion the melting temperature $T_m$ is related to the Debye temperature $\Theta_D$ as follows: Assuming a Debye solid, $T_m = C\, V^{2/3}\, \Theta_D^2$, where V is the volume and C is a constant to be derived for each specific metal. Assuming that $\gamma = \gamma_o\, (\rho_o/\rho)^q$ and $q = 1$ one gets the approximation:

$$T_m(V) = T_{mo}\, (V/V_o)^{2/3}\, \exp[2\,\gamma_o\,(1 - V/V_o)] \qquad (2)$$

The third-order Birch–Murnaghan (BM) equation of state
$$P(V) = 3/2\, B_o\, [\,(V_o/V)^{7/3} - (V_o/V)^{5/3}\,]\,[1 + 3/4\,(B_o' - 4)\{(V_o/V)^{2/3} - 1\}] \qquad (3)$$

We imply the constrain that the fitting parameter of the experimental equation of state (EOS) at ambient temperature must simultaneously fit the experimental melting data for deriving $\gamma_o$.

For the reader who wants to use the combined approach procedure using the MATLAB programs:

## VINET (VIN) EOS combined with Lindemann-Gilvarry melting approximation:

```
B=B₀
dB=B'
V=Vo:-0.086:4     (Vo is the Volume at RT)
VVo=V/Vo;
A=1.5*B*(VVo.^-2.333)-(VVo.^-1.666);
C=1-(0.75*(4-dB));
D=C*((VVo.^-0.6666)-1);
E=D.*C;;
P=E.*A;
r=VVo;
Gama= γ₀
EX=exp(2*Gama*(1-r));
Tm=TM₀*r.^0.6666;
TM=EX.*Tm;
plot(P,TM,'r')
plot(P,V,'b')
```

## Third order Birch-Murnaghan (BM) EOS combined with Lindenmann-Gilvarry melting approximation:

```
B=B₀
dB=B'
V=Vo:-0.086:4
Vo=(Vo is the Volume at RT)
VVo=V/Vo;
A=1.5*B*(VVo.^-2.333)-(VVo.^-1.666);
C=1-(0.75*(4-dB));
D=C*((VVo.^-0.6666)-1);
E=D.*C;;
P=E.*A;
r=VVo;
Gama=γ₀
EX=exp(2*Gama*(1-r));
Tm=TM₀*r.^0.6666;
TM=EX.*Tm;
plot(P,TM,'r')
plot(P,V,'b')
```

Here the bulk moduli $B=B_o$, $dB/dP=B_o'$ and Gama= $\gamma_o$ is the lattice Grüneisen parameter. $\gamma$ is a fitting parameter. TM is the melting curve according to the Gilvarry-Lindemann criterion.

## 2. Lindemann-Gilvarry vs. Simon-Glatzel fitting curve procedure

The Simon-Glatzel equation is in fact combination of Murnghan EOS and Lindemann's criterion: $Tm = T_{ref}(B_o'(P-P_{ref}/B_o+1)^{2(\gamma-1/3+f)}$, here f is the coefficient in Mores potential.

The advantage of the constraint Gilvarry-Lindeman procedure used in the present contribution, is that the volume, the bulk moduli and $\gamma_o$ are simultaneously derived.